\documentclass{article}

\usepackage{graphicx}
\usepackage{amsmath}
\author{Choulakian, V.,  Universit\'{e} de Moncton, Canada ;\\ vartan.choulakian@umoncton.ca}
\title{Taxicab Correspondence Analysis of Sparse Two-Way Contingency Tables;}

\begin{document}
\maketitle

\begin{abstract}
Visualization and interpretation of contingency tables by correspondence
analysis (CA), as developed by Benz\'{e}cri, have a rich structure based on
Euclidean geometry. However, it is a well established fact that, often CA is
very sensitive to sparse contingency tables, where we characterize sparsity
as the existence of relatively high-valued counts, rare observations and
zero-block structure. Our main aim in this paper is to highlight the above
mentioned three interrelated points in two ways: First, we propose a
7-number summary of sparsity based on the minimal size of an equivalent
contingency table, where the invariance property of CA and TCA (a L$_{1}$
variant of CA named taxicab) is used to construct the equivalence class of
contingency tables; second, we compare the maps obtained by CA and TCA to
explore under what conditions the CA and TCA maps produce similar, somewhat
similar or dissimilar maps. Examples are provided.

Key words: Sparse contingency tables; correspondence analysis; taxicab
correspondence analysis; interpretable maps; 7-number summary of sparsity.
\end{abstract}

\section{\bf Introduction}

Correspondence analysis (CA), developed by Benz\'{e}cri (1973) since 1960s,
as a statistical method for different kinds of data sets, in particular for
contingency tables, is embedded both in theory and in practice. The theory
is based on the chi-square distance between the profiles; parallel to this
beautiful theory, the practice is entrenched in the joint interpretation of
the graphical displays based on the Euclidean geometry. Seeing this extreme
fondness of the use and interpretation of the maps by the users of CA,
Nishisato (1998) suggested the replacement of the adage \textquotedblleft
seeing is believing\textquotedblleft\ with \textquotedblleft graphing is
believing\textquotedblleft\ and stressed the importance of interpretable
graphs. Additionally, we recall the often cited quip \textquotedblleft a
picture is worth a thousand words\textquotedblleft , and via geometric\
interpretation of maps CA offers much to the analysis of complex
multivariate data sets. So the philosophical question asked by Schlick
(2000, part 5) "Theory and Observation: Is seeing believing ?" is quite
relevant here in the context of data analysis by CA.

It is well known that, CA is very sensitive to some particularities of a
data set; further, how to identify and handle these is an open unresolved
problem. Here, we enumerate three under the umbrella of sparse contingency
tables: rare observations, zero-block structure and relatively high-valued
cells. Rao (1995), among others, stressed the influence of rare observations
(rows or columns that have relatively small marginal weights compared to
others) and proposed an alternative to CA based on Hellinger distance ( a
square-root transformation of counts). Greenacre (2013) refuted Rao's
assertion and argued that rare observations do not have an exaggerated
influence in CA. Earlier Nowak and Bar-Hen (2005) developed a criterion
based on the influence function to identify influential rare observations;
and they arrived at the same conclusion as Greenacre in their analysis of a $%
207\times 15$ abundance data in ecology; however they observed that
"influential species are rare species that are concentrated in few plots". A
similar observation is found in Greenacre (2013) "there is one exceptional
situation where rare species would have a strong role in the solution,
namely when a species is observed in a single sampling site and no or very
few other species are observed there". We describe this particular situation
as the existence of a large zero-block structure. Often few relatively
high-valued cells, including outlier counts, have detrimental effect on the
CA outputs by emphasizing some aspects of the data, even though apparently
the interpretation of the CA maps seems meaningful to the researchers. Our
main aim in this paper is to highlight the above mentioned three points by
comparing the maps obtained by CA with the maps obtained by taxicab
correspondence analysis (TCA), where TCA is a L$_{1}$ variant of CA; and to
explore under what conditions the CA and TCA maps produce similar, somewhat
similar or dissimilar maps. Our main conclusion is that: First, CA and TCA
maps enrich each other; second, for sparse contingency tables, there is a
positive probability that CA and TCA maps are partially similar or
dissimalar. To do this we organize the paper in six sections.

In section 2, we attempt to quantify the notion of sparsity in contingency
tables by a 7-number summary based on the minimal size of an equivalent
contingency table, where the invariance property of CA and TCA is used to
construct the equivalence class of contingency tables. In section 3, we
present a brief mathematical comparison of CA and TCA; in section 4 we
present an empirical comparison using ten data sets; in section 5 we
consider sparsest contingency tables; and we conclude in section 6.

The theory of CA can be found, among others, in Benz\'{e}cri (1973, 1992),
Greenacre (1984), Gifi (1990), Le Roux and Rouanet (2004), Murtagh (2005),
and Nishisato (2007); the recent book, authored by Beh and Lombardi (2014),
presents a panoramic review of CA and related methods. Since 2006,
Choulakian and coauthors have studied mathematical properties of TCA applied
to many kinds of non-negative data; in particular, TCA of contingency tables
and their comparison with CA are studied in the following papers: Choulakian
(2006), Choulakian et al. (2006), Choulakian (2008), and Choulakian,
Simonetti and Gia (2014).

\section{7-number summary of sparsity in contingency tables}

Let ${\bf N=(}n_{ij})$ be a contingency table cross-classifying two nominal
variables with $I$ rows and $J$ columns, where for $i=1,...,I$ and $%
j=1,...,J,$ $n_{ij}$ represents the frequency of statistical units having
the $i$th category of the row variable and the $j$th category of the column
variable. Thus, $n=\sum_{i=1}^{I}\sum_{j=1}^{J}n_{ij}$ represents the sample
size. In the statistical literature, generally we see that the degree of
sparsity of {\bf N} are based on the following two quantities
\[
ave({\bf N})=\frac{\sum_{i=1}^{I}\sum_{j=1}^{J}n_{ij}}{IJ},
\]%
the average value of counts; and
\[
\%(0\in {\bf N})=\frac{\sum_{i=1}^{I}\sum_{j=1}^{J}1_{n_{ij}=0}}{IJ}100,
\]%
the percentage value of zero counts, where $1_{n_{ij}=0}$ is the indicator
function: $1_{n_{ij}=0}=1$ for $n_{ij}=0$ and $1_{n_{ij}\neq 0}=0$ for $%
n_{ij}\geq 1.$

According to Agresti and Yang (1987), ${\bf N}$ is {\it sparse} if $ave({\bf %
N})$ is small such that the chi-squared approximations of the
goodness-of-fit statistics are inaccurate. Radavicius and Samusenko (2012)
characterize ${\bf N}$ as {\it very sparse} if the sample size ($n$) is less
than the number of cells ($IJ$), that is, $ave({\bf N})<1$. Greenacre (2013)
uses $\%(0\in {\bf N})$ as an index of sparsity.

Another qualitative definition of sparsity is used in the Ph.D thesis of
Kraus (2012), based on Agresti (2002, p.391) "contingency tables having
small cell counts are said to be sparse". A quantification of this
definition will be given in subsection 2.3.

As we stated in the introduction, our concept of sparseness is broader, it
also includes relatively large valued counts; to quantify this aspect of
sparseness we consider the batch of nonzero counts of ${\bf N}$, and
following Tukey (1977, ch.2 or p.80), we summarize them by the 5-number
summary, $$MH1=(min,Q1,Median,Q3,max);$$ where, $min$ represents the lowest
value in the batch of the positive counts, $max$ the highest value, and, $Q1$%
, $Median$ and $Q3$ are the three quartiles ($Q1$ and $Q3$ are the two
hinges in Tukey's terminology). Thus, from the 7-number summary ($ave({\bf X}%
),\%(0\in {\bf X}),MH1),$ one gets an idea on the degree of sparsity
concerning its different, but complementary, aspects in a contingency table $%
{\bf X}$.
\begin{verbatim}
2.1 Equivalence class of an observed contingency table
\end{verbatim}

An important property of CA and TCA is that columns or rows with identical
profiles (conditional probabilities) receive identical factor scores. The
factor scores are used in the graphical displays. Moreover, merging of
identical profiles does not change the results of the data analysis: This is
named the {\it principle of equivalent partitioning} by Nishisato (1984); it
includes the famous invariance property named {\it principle of
distributional equivalence}, on which Benz\'{e}cri (1973) developed CA.
Formally, Nishisato's {\it principle of equivalent partitioning }is based on
the following \bigskip

{\bf Definition 1}: Let ${\bf N}$ be a contingency table of
size $I\times J$,  ${\bf x=(}x_{k})$ and  ${\bf y=(}y_{k})$ are two rows or two columns of ${\bf N}$  such that they are proportional
\[
\frac{{\bf x}}{\sum x_{i}}=\frac{{\bf y}}{{\bf \sum }y_{i}}\text{{\bf \ \ \
\ }or{\bf \ \ \ }}{\bf (\sum }y_{i}){\bf x=\ (}\sum x_{i}){\bf y}.
\]%

We construct a new contingency table, $N_{reduced}$, by replacing the two elements ${\bf x}$ and  ${\bf y}$ in ${\bf N}$ by one element 
${\bf x}+{\bf y}$, and keeping all the other columns and rows of ${\bf N}$ the same in $N_{reduced}$. Then we say that the contingency tables ${\bf N}$ and $N_{reduced}$ are equivalent, and we write ${\bf N}\sim N_{reduced}$.\\
Thus the equivalence class of contingency tables of ${\bf N}$ is
given by%
\[
\Omega ({\bf N)=}\left\{ {\bf X}:{\bf X}\sim {\bf N}\right\} .
\]%
Given that, $\Omega ({\bf N)}$ contains infinite number of contingency
tables equivalent to a given ${\bf N}$, we define its representative element
by the unique contingency table ${\bf M}$ of minimal size; that is, among
all elements of $\Omega ({\bf N)}$, ${\bf M}$ has minimum number of rows and
columns. We can easily deduce the following inequalities: $ave({\bf N})\leq
ave({\bf M})$ and $\max ({\bf N})\leq \max ({\bf M}).$
\begin{verbatim}
2.2 Artificial example and extreme sparsity
\end{verbatim}

The following contrived example illustrates the idea. Let
\[
{\bf N}=\left(
\begin{array}{cccc}
1 & 2 & 0 & 0 \\
2 & 4 & 0 & 0 \\
0 & 0 & 1 & 2 \\
3 & 6 & 0 & 0%
\end{array}%
\right)
\]%
be a two-way contingency table of size $4\times 4$. Its 7-number summary of
sparsity is
\[
(ave({\bf N})=1.3125,\ \%(0\in {\bf N})=50,MH1=(1,\ 1.5,\ 2,\ 3.5,\ 6)).
\]%
We note that the first, second and fourth rows of ${\bf N}$ are proportional
to each other, so they can be lumped together into one row, and we obtain
the equivalent contingency table
\[
{\bf N}_{1}=\left(
\begin{array}{cccc}
6 & 12 & 0 & 0 \\
0 & 0 & 1 & 2%
\end{array}%
\right)
\]%
of size $2\times 4$; its 7-number summary of sparsity is $(2.6250,\ 50,\
(1,\ 1.5,\ 4,\ 9,\ 12))$. Similarly, we see that the third and fourth
columns of ${\bf N}_{1}$ are proportional, so they can be added together,
and we obtain equivalent contingency table
\[
{\bf N}_{2}=\left(
\begin{array}{ccc}
6 & 12 & 0 \\
0 & 0 & 3%
\end{array}%
\right)
\]%
of size $2\times 3$. Similarly, we see that the first and second columns of $%
{\bf N}_{2}$ are proportional, so they can be added together, and we obtain
the unique representative equivalent contingency table%
\[
{\bf M}=\left(
\begin{array}{cc}
18 & 0 \\
0 & 3%
\end{array}%
\right)
\]%
of minimal size $2\times 2.$ The four contingency tables ${\bf N}$, ${\bf N}%
_{1},$ ${\bf N}_{2}$ and ${\bf M}$ are equivalent, because they belong to $%
\Omega ({\bf N):}$ CA and TCA of ${\bf N}$, ${\bf N}_{1},$ ${\bf N}_{2}$ and
${\bf M}${\bf \ }produce identical maps, because they have identical
geometries within the mathematical framework of CA and TCA. However, we have
four different 7-number summaries of the sparsity: we consider the one
obtained from ${\bf M}$ the most representative.

We note that {\bf M} is a diagonal contingency table and sparsest (most
sparse), based on the following lemma, whose proof is given in the
appendix.\bigskip

{\bf Lemma 1}: $\%(0\in {\bf M})\leq 100(1-\frac{1}{\min (I,J)}).\bigskip $

{\bf Definition 2}: A contingency table is named {\it sparsest} if $\%(0\in
{\bf M})=100(1-\frac{1}{\min (I,J)}),$ and extremely sparse if $\%(0\in {\bf %
M})$ is very near to $100(1-\frac{1}{\min (I,J)}).$
\begin{verbatim}
2.3 Examples of sparse contingency data sets
\end{verbatim}

Table 1 enumerates ten contingency tables and their 7-number summaries
calculated on ${\bf N}$ and on ${\bf M}$. Sections 4 and 5 provide further
references to these data sets. The first data set is not sparse. For the
last nine of them, which are considered to be sparse, we note that:%
\[
Q1\leq 2\ \ \ \ \ \text{and\ \ \ \ \ }Median\leq 5,
\]%
which is another quantification of sparsity describing "contingency tables
having small cell counts are said to be sparse". Furthermore, comparison of $%
Q3$ and $max$ values highlights very long tails for sparse contingency
tables, which represents the existence of relatively high-valued counts.
Concerning the equivalent tables ${\bf N}$ and ${\bf M}${\bf , }we see
noticeable changes in the 7-number summaries for the two data sets 6 ({\it %
Barents}) and 10 ({\it Synoptic Gospels}): these two contingency tables $%
{\bf N}$ and ${\bf M}${\bf \ }are extremely tall: the number of columns is
much smaller than the number of rows; so the merging of rows essentially
happened for rows having very small marginal counts of 1 or 2. For these two
data sets, ${\bf M}$ can be put in the following form%
\[
{\bf M}=\left(
\begin{array}{c}
{\bf M}_{1} \\
{\bf D}%
\end{array}%
\right) ,
\]%
where ${\bf D}$ is a square diagonal matrix.

We classify the data sets in Table 1 into three large groups according to
our concept of sparsity:

Non sparse tables: Data set 1 ({\it TV programs}){\it \ }belongs to this
group.

Extremely sparse tables: Data set 3 ({\it Texel}) belongs to this group.
Note that $\%(0\in {\bf M})=96.3\%$ is very near to $100(1-1/220)=99.5455,$
the upper bound provided in Lemma 1.

Sparse tables: the remaining eight data sets belong to this group.

It is interesting to note that for Data set 10 ({\it Synoptic Gospels)}: The
upper bound in Lemma 1, $100(1-1/7)=85.7143$, is quite near to $\%(0\in {\bf %
N})=78.2\%$, but quite far from $\%(0\in {\bf M})=45\%.\ $For this reason,
we characterized it as sparse and not extremely sparse.

\bigskip
{\tiny{
\begin{tabular}{|l|l|l|l|lllll|l|}
\multicolumn{10}{l|}{\bf Table 1: 7-number summary of sparsity of ten
two-way contingency tables.} \\ \hline
& size & ave & \%(0) & MH1 & \multicolumn{1}{l}{} & \multicolumn{1}{l}{} &
\multicolumn{1}{l}{} & \multicolumn{1}{l|}{} & map \\
& &  &  &  & \multicolumn{1}{l}{} & \multicolumn{1}{l}{} &
\multicolumn{1}{l}{} & \multicolumn{1}{l|}{} &similarity \\ \hline
1) {\it TV } &  &  & \multicolumn{1}{|l|}{} & \multicolumn{1}{|l}{}
& \multicolumn{1}{l}{} & \multicolumn{1}{l}{} & \multicolumn{1}{l}{} &
\multicolumn{1}{l|}{} & yes \\
{\it programs} &  &  & \multicolumn{1}{|l|}{} & \multicolumn{1}{|l}{}
& \multicolumn{1}{l}{} & \multicolumn{1}{l}{} & \multicolumn{1}{l}{} &
\multicolumn{1}{l|}{} & \\ \hline

{\bf N=M} & $13\times 7$ & 55.81 & 0\% & (3 & 15 & 40 & 86 & 271) &  \\
\hline
2) {\it Rodents} &  &  &  &  &  &  &  &  & no \\ \hline
{\bf N} & $28\times 9$ & 3.96 & 66.7\% & (1 & 2 & 5 & 12.3 & 78) &  \\
{\bf M} & $21\times 9$ & 5.3 & 58.7\% & (1 & 2 & 4.5 & 14 & 78) &  \\ \hline
3) {\it Texel} &  &  &  &  &  &  &  &  & no \\ \hline
{\bf N} & $285\times 220$ & 0.26 & 96.6\% & (1 & 1 & 1 & 4.8 & 97) &  \\
{\bf M} & $266\times 220$ & 0.28 & 96.3\% & (1 & 1 & 1 & 7 & 97) &  \\ \hline
4) {\it Macro} &  &  &  &  &  &  &  &  & partial \\ \hline
{\bf N} & $189\times 40$ & 6.1 & 84.8\% & (1 & 2 & 3 & 14 & 1848) &  \\
{\bf M} & $161\times 40$ & 7.47 & 81.9\% & (1 & 2 & 3 & 14 & 1848) &  \\
\hline
5) {\it Benthos} &  &  &  &  &  &  &  &  & partial \\ \hline
{\bf N=M} & $92\times 13$ & 8.02 & 39\% & (1 & 1 & 3 & 8 & 992) &  \\ \hline
6) {\it Barents} &  &  &  &  &  &  &  &  & partial \\ \hline
{\bf N} & $446\times 10$ & 2.91 & 78.4\% & (1 & 1 & 2 & 8 & 798) &  \\
{\bf M} & $221\times 10$ & 5.87 & 67.5\% & (1 & 1 & 3 & 10 & 903) &  \\
\hline
7) {\it Seashore} &  &  &  &  &  &  &  &  & partial \\ \hline
{\bf N} & $126\times 68$ & 0.14 & 88\% & (1 & 1 & 1 & 1 & 5) &  \\
{\bf M} & $106\times 65$ & 0.17 & 86.4\% & (1 & 1 & 1 & 1 & 12) &  \\ \hline
8) {\it Punta } &  &  &  &  &  &  &  &  & no \\
{\it  Milazzese} &  &  &  &  &  &  &  &  &  \\ \hline
{\bf N=M} & $31\times 19$ & 0.83 & 58.1\% & (1 & 1 & 1 & 2 & 12) &  \\ \hline
9) {\it Iversfjord} &  &  &  &  &  &  &  &  & no \\ \hline
{\bf N=M} & $37\times 14$ & 2.643 & 60\% & (1 & 1 & 2 & 6 & 64) &  \\ \hline
10) {\it Synoptic } &  &  &  &  &  &  &  &  & no \\
{\it  Gospels} &  &  &  &  &  &  &  &  &  \\ \hline
{\bf N} & $7097\times 7$ & 0.39 & 78.2\% & (1 & 1 & 1 & 2 & 79) &  \\
{\bf M} & $796\times 7$ & 3.59 & 45\% & (1 & 1 & 2 & 4 & 2740) &  \\ \hline
\end{tabular}}
}
\section{Correspondence analysis and taxicab correspondence analysis: an
overview}

Let ${\bf P=N/}n=(p_{ij})$ be the associated correspondence matrix of {\bf N}%
. We define as usual $p_{i\ast }=\sum_{j=1}^{J}p_{ij}$ , $p_{\ast
j}=\sum_{i=1}^{I}p_{ij},$ the vector ${\bf r=(}p_{i\ast })\in
\mathbf{R}
^{I},$ the vector ${\bf c=(}p_{\ast j})\in
\bf{R}
^{J}$, and ${\bf D}_{r}=Diag({\bf r})$ the diagonal matrix having diagonal
elements $p_{i\ast },$ and similarly ${\bf D}_{c}=Diag({\bf c}).$ We suppose
that ${\bf D}_{r}$ and ${\bf D}_{c}$ are positive definite metric matrices
of size $I\times I$ and $J\times J$, respectively; this means that the
diagonal elements of ${\bf D}_{r}$ and ${\bf D}_{c}$ are strictly positive.
Let $k=rank({\bf R}_{0}{\bf ),}$ where
\[
{\bf R}_{0}=({\bf P}-{\bf rc}^{\top })
\]%
is the residual matrix with respect to the independence model. CA and TCA
can be considered as principal components analysis for categorical data,
where ${\bf P}$ or ${\bf R}_{0}$ is decomposed into a sum of bilinear terms
shown in equation (1). Equation (1) is named the data reconstruction
formula, and it is obtained by generalized singular value decomposition and
its taxicab version with respect to the metric matrices ${\bf D}_{r}$ and $%
{\bf D}_{c}$, see in particular Choulakian, Simonetti and Gia (2014):

\[
{\bf P}={\bf D}_{r}({\bf 1}_{I}{\bf 1}_{J}^{^{\top }}+\sum_{\alpha =1}^{k}%
{\bf f}_{\alpha }{\bf g}_{\alpha }^{^{\top }}/\sigma _{\alpha }){\bf D}_{c},
\]%
or elementwise%
\begin{equation}
p_{ij}=p_{i\ast }p_{\ast j}\left[ 1+\sum_{\alpha =1}^{k}f_{\alpha
}(i)g_{\alpha }(j)/\sigma _{\alpha }\right] ,
\end{equation}%
where ${\bf f}_{\alpha }$\ and ${\bf g}_{\alpha }$ represent the principal
coordinate scores of rows and columns, and $\sigma _{\alpha }$ is the
associated dispersion measure for\ \ $\alpha =1,...,k.$ Note that in both
methods ${\bf f}_{\alpha }$\ and ${\bf g}_{\alpha }$ are ${\bf D}_{r}$ and $%
{\bf D}_{c}$ centered respectively; that is
\begin{eqnarray}
{\bf f}_{\alpha }^{^{\top }}{\bf D}_{r}{\bf 1}_{I} &=&{\bf g}_{\alpha
}^{^{\top }}{\bf D}_{c}{\bf 1}_{J}  \nonumber \\
&=&0,
\end{eqnarray}%
where ${\bf 1}_{I}$ is a column vector of ones of size $I$.

In CA, ${\bf f}_{\alpha }$\ and ${\bf g}_{\alpha }$ satisfy%
\begin{equation}
{\bf f}_{\alpha }^{^{\top }}{\bf D}_{r}{\bf f}_{\alpha }={\bf g}_{\alpha
}^{^{\top }}{\bf D}_{c}{\bf g}_{\alpha }=\sigma _{\alpha }^{2}\text{ \ \
for\ \ }\alpha =1,...,k,
\end{equation}%
\begin{equation}
{\bf f}_{\alpha }^{^{\top }}{\bf D}_{r}{\bf f}_{\beta }={\bf g}_{\alpha
}^{^{\top }}{\bf D}_{c}{\bf g}_{\beta }=0\text{ \ \ for\ \ }\alpha \neq
\beta .
\end{equation}%
Equation (3) says that the ${\bf D}_{r}$ weighted L$_{2}$ norm of ${\bf f}%
_{\alpha }$ is $\sigma _{\alpha };$ likewise, equation (4) says that ${\bf f}%
_{\alpha }$ is ${\bf D}_{r}$ orthogonal to ${\bf f}_{\beta }$ \ for\ \ $%
\alpha \neq \beta $. In CA the standard coordinate scores are ${\bf f}%
_{\alpha }/\sigma _{\alpha }$ for column profiles and ${\bf g}_{\alpha
}/\sigma _{\alpha }$ for row profiles.

In TCA, ${\bf f}_{\alpha }$\ and ${\bf g}_{\alpha }$ satisfy%
\begin{equation}
{\bf f}_{\alpha }^{^{\top }}{\bf D}_{r}sgn({\bf f)}_{\alpha }={\bf g}%
_{\alpha }^{^{\top }}{\bf D}_{c}sgn({\bf g}_{\alpha })=\sigma _{\alpha }%
\text{ \ \ for\ \ }\alpha =1,...,k,
\end{equation}%
\begin{equation}
{\bf f}_{\alpha }^{^{\top }}{\bf D}_{r}sgn({\bf f}_{\beta })={\bf g}_{\alpha
}^{^{\top }}{\bf D}_{c}sgn({\bf g)}_{\beta }=0\text{ \ \ for\ \ }\alpha
>\beta .
\end{equation}%
where $sgn({\bf g}_{\alpha })=[sgn(g_{\alpha }(1)),...,sgn(g_{\alpha
}(J)]^{^{\top }},$ and $sgn(g_{\alpha }(j))=1$ if $g_{\alpha }(j)>0,$ $%
sgn(g_{\alpha }(j))=-1$ otherwise. Equation (5) says that the ${\bf D}_{r}$
weighted $L_{1}$ norm of ${\bf f}_{\alpha }$ is $\sigma _{\alpha };$
likewise, equation (6) says that ${\bf f}_{\alpha }$ is ${\bf D}_{r}$
orthogonal to $sgn{\bf (f}_{\beta })$ \ for\ \ $\alpha >\beta $.
\begin{verbatim}
2.1: Remarks
\end{verbatim}

\begin{itemize}
\item a) CA of ${\bf P}${\bf \ }is equivalent to CA of ${\bf R}_{0}$, with
diagonal weight matrices ${\bf D}_{r}$ and ${\bf D}_{c}.$ Analogously, TCA
of ${\bf P}${\bf \ }is equivalent to TCA of ${\bf R}_{0},$ with diagonal
weight matrices ${\bf D}_{r}$ and ${\bf D}_{c}.$

\item b) In CA, the principal coordinate scores ${\bf f}_{\alpha }$\ and $%
{\bf g}_{\alpha }$ are functions of the eigenvectors of a similarity measure
between the rows or columns and more importantly the similarity measure
depends on the chosen metric ${\bf D}_{r}$ and ${\bf D}_{c}.$ We describe
the computation of ${\bf f}_{\alpha }$ and ${\bf g}_{\alpha }$ in four steps:

\ \ \ Step 1: we calculate the matrix of Pearson residuals,%
\begin{equation}
{\bf S}={\bf D}_{r}^{-1/2}({\bf P}-{\bf rc}^{^{\top }}){\bf D}_{c}^{-1/2}.
\end{equation}

\ \ \ Step 2: we calculate the eigenvectors ${\bf x}_{\alpha }$ via the
eigen-equation,%
\begin{equation}
{\bf S}^{^{\top }}{\bf Sx}_{\alpha }=\sigma _{\alpha }^{2}{\bf x}_{\alpha }%
\text{ \ \ \ with \ }{\bf x}_{\alpha }^{^{\top }}{\bf x}_{\alpha }=1,
\end{equation}%
where the $(i,j)$th element of ${\bf S}^{^{\top }}{\bf S}$ represents a
similarity measure between the two column categories $i$ and $j$.

\ \ \ Step 3: we calculate ${\bf f}_{\alpha }=\sigma _{\alpha }{\bf D}%
_{r}^{-1/2}{\bf x}_{\alpha }.$

\ \ \ Step 4: we calculate ${\bf g}_{\alpha }$ via the transition formula
(22).

\item c) Compared to CA, TCA stays as close as possible to the original
data: It directly acts on the correspondence matrix ${\bf P}${\bf \ }or{\bf %
\ }${\bf R}_{0}$ in the largest sense that the basic taxicab decomposition
is independent of the metrics ${\bf D}_{r}$ and ${\bf D}_{c}$: it is simply
constructed from a sum of the signed columns or rows of the residual
correspondence matrix, for further details see Choulakian (2006, 2016); only
the relative direction of the rows or columns is taken into account without
calculating a similarity (or dissimilarity) measure between the rows or
columns.

The optimization criterion is based on the famous Grothendieck problem, see
Pisier \ (2012). The steps for the computation of the principal coordinate
scores ${\bf f}_{\alpha }$ and ${\bf g}_{\alpha }$ are done iteratively for$%
\ \alpha =1,...,k:$
\end{itemize}

Step 1: we compute the principal axis%
\[
{\bf u}_{\alpha }=\arg \max_{{\bf u\in }\left\{ -1,1\right\} ^{J}}||{\bf R}%
_{\alpha -1}{\bf u}||_{1},
\]%
where ${\bf R}_{0}={\bf P}-{\bf rc}^{^{\top }}$ $\ $and $\ {\bf R}_{\alpha }=%
{\bf P}-{\bf rc}^{^{\top }}-\sum_{\beta =1}^{\alpha }{\bf D}_{r}{\bf f}%
_{\beta }{\bf g}_{\beta }^{^{\top }}{\bf D}_{c}/\sigma _{\beta }$ for \ $%
\alpha =1,...,k$.

Step 2: we compute the principal coordinate scores ${\bf f}_{\alpha }={\bf D}%
_{r}^{-1}{\bf R}_{\alpha -1}{\bf u}_{\alpha }$.

Step 3: we calculate ${\bf g}_{\alpha }$ via the transition formula (15), $$%
{\bf g}_{\alpha }={\bf D}_{c}^{-1}{\bf R}_{\alpha -1}^{^{\top }}sgn({\bf f}%
_{\alpha })$$.

Step 4: we update ${\bf R}_{\alpha +1}={\bf P}-{\bf rc}^{^{\top
}}-\sum_{\beta =1}^{\alpha +1}{\bf D}_{r}{\bf f}_{\beta }{\bf g}_{\beta
}^{^{\top }}{\bf D}_{c}/\sigma _{\beta }.$

\begin{itemize}
\item d) An interesting and useful property of the taxicab dispersion
measures, $\sigma _{\alpha }$ for $\alpha \geq 1,$ is the following result
well known in theoretical computer science, see Khot and Naor (2013):

{\bf Lemma 2}:
\begin{eqnarray*}
\sigma _{\alpha } &=&\frac{||{\bf R}_{\alpha -1}{\bf u}||_{1}}{||{\bf u}%
||_{\infty }}\ \ \ \ \ \text{for \ \ \ }\alpha \geq 1 \\
&=&\max_{{\bf u\in }\left\{ -1,1\right\} ^{J}}||{\bf R}_{\alpha -1}{\bf u}%
||_{1} \\
&=&4\ ||{\bf R}_{\alpha -1}||_{cut},
\end{eqnarray*}%
where the cut norm of the matrix ${\bf R}_{\alpha -1}$ is defined as
\[
||{\bf R}_{\alpha -1}||_{cut}=\max_{S\times T}\ |\sum_{(i,j)\in S\times T}%
{\bf R}_{\alpha -1}(i,j)|\text{\ \ where\ }\\S\subseteq \left\{
1,...,I\right\}\]
\[\text{ and }T\subseteq \left\{ 1,...,J\right\} .
\]%
We know that taxicab principal axes have values $\pm 1,$ that is,$\ {\bf u}%
_{\alpha }{\bf \in }\left\{ -1,1\right\} ^{J}$ and ${\bf v}_{\alpha }{\bf %
\in }\left\{ -1,1\right\} ^{I}$ for $\alpha \geq 1$. So we can represent $%
{\bf u}_{\alpha }={\bf u}_{\alpha +}+{\bf u}_{\alpha -},$ and similarly, $%
{\bf v}_{\alpha }={\bf v}_{\alpha +}+{\bf v}_{\alpha -}$, where
\begin{eqnarray*}
{\bf u}_{\alpha +} &=&({\bf 1}_{J}+{\bf u}_{\alpha })/2 \\
{\bf u}_{\alpha -} &=&({\bf u}_{\alpha }-{\bf 1}_{J})/2.
\end{eqnarray*}%
Lemma 2 can be named 4-quadrants balancing property, because the taxicab
dispersion measure $\sigma _{\alpha }$ for $\alpha \geq 1$ is divided into 4
equal parts having the common value of the cut norm of ${\bf R}_{\alpha -1}$%
:
\begin{eqnarray*}
\sigma _{\alpha }/4 &=&{\bf v}_{\alpha +}^{\prime }{\bf R}_{\alpha -1}{\bf u}%
_{\alpha +} \\
&{\bf =}&{\bf v}_{\alpha -}^{\prime }{\bf R}_{\alpha -1}{\bf u}_{\alpha -} \\
&=&|{\bf v}_{\alpha -}^{\prime }{\bf R}_{\alpha -1}{\bf u}_{\alpha +}| \\
&=&|{\bf v}_{\alpha +}^{\prime }{\bf R}_{\alpha -1}{\bf u}_{\alpha -}|.
\end{eqnarray*}%
As a corollary to this fact, we have: In TCA of ${\bf P}$ both principal
coordinate scores ${\bf f}_{\alpha }$ and ${\bf g}_{\alpha }$ for$\ \ \alpha
=1,...,k$ satisfy the equivariability property, see Choulakian (2008b). This
means that ${\bf f}_{\alpha }$ and ${\bf g}_{\alpha }$ are equally balanced
in the sense that%
\begin{eqnarray}
\frac{\sigma _{\alpha }}{2} &=&\sum_{i\in I_{\alpha +}}p_{i\ast }f_{\alpha
}(i),  \nonumber \\
&=&-\sum_{i\in I_{\alpha -}}p_{i\ast }f_{\alpha }(i),  \nonumber \\
&=&\sum_{j\in J_{\alpha +}}p_{\ast j}g_{\alpha }(j),  \nonumber \\
&=&-\sum_{j\in J_{\alpha -}}p_{\ast j}g_{\alpha }(j),
\end{eqnarray}%
where $I_{\alpha +}=\left\{ i|f_{\alpha }(i)>0\right\} $, $I_{\alpha
-}=\left\{ i|f_{\alpha }(i)<0\right\} ,$ $J_{\alpha +}=\left\{ j|g_{\alpha
}(j)>0\right\} $ and $J_{\alpha -}=\left\{ j|g_{\alpha }(j)<0\right\} .$
This easily follows from the fact that the principal coordinate scores ${\bf %
f}_{\alpha }$ and ${\bf g}_{\alpha }$ are ${\bf D}_{r}$ and ${\bf D}_{c}$
centered, they satisfy equation (2). An informal illustrative interpretation
of the equivariability property is that TCA pulls inside potential
influential observations and pushes outside points around the origin, thus
providing a more balanced and robust view of data.
\end{itemize}

We note that in CA the principal coordinate scores ${\bf f}_{\alpha }$ and $%
{\bf g}_{\alpha }$ do not satisfy the equivariability property, because they
are unequally balanced in the sense that%
\begin{eqnarray*}
A &=&\sum_{i\in I_{\alpha +}}p_{i\ast }f_{\alpha }(i), \\
&=&-\sum_{i\in I_{\alpha -}}p_{i\ast }f_{\alpha }(i), \\
B &=&\sum_{j\in J_{\alpha +}}p_{\ast j}g_{\alpha }(j), \\
&=&-\sum_{j\in J_{\alpha -}}p_{\ast j}g_{\alpha }(j),
\end{eqnarray*}%
and in general,
\[
A\neq B;
\]%
furthermore, $A$ and $B$ are not related to the dispersion measure $\sigma
_{\alpha },$ because CA maximizes the variance of the principal coordinate
scores.

\begin{itemize}
\item e) Given that the approach in CA and TCA is geometric, influence
measure of a point (a column or a row) to the $\alpha $th factor is provided
by the contribution of that point to the dispersion measure of the $\alpha $%
th factor in per 1000 units.

In CA, based on (3), this corresponds to:%
\begin{equation}
C_{\alpha }(i)=1000\frac{p_{i\ast }f_{\alpha }^{2}(i)}{\sigma _{\alpha }^{2}}%
\text{ \ \ and \ \ \ }C_{\alpha }(j)=1000\frac{p_{j\ast }g_{\alpha }^{2}(j)}{%
\sigma _{\alpha }^{2}}\text{.}
\end{equation}%
In TCA, based on (5), we have the signed contribution%
\begin{equation}
SC_{\alpha }(i)=1000\frac{p_{i\ast }f_{\alpha }(i)}{\sigma _{\alpha }}\text{
\ \ and \ \ \ }SC_{\alpha }(j)=1000\frac{p_{j\ast }g_{\alpha }(j)}{\sigma
_{\alpha }}\text{.}
\end{equation}%
It is important to note that, in CA,
\begin{equation}
0<C_{\alpha }(point)<1000\text{;}
\end{equation}%
while in TCA, from (9) we get,%
\begin{equation}
-500\leq SC_{\alpha }(point)\leq 500.
\end{equation}

\item f) In both methods the maps or joint displays are obtained by plotting
(${\bf f}_{\alpha },{\bf f}_{\beta })$\ and (${\bf g}_{\alpha },{\bf g}%
_{\beta })$ for $\alpha \neq \beta .$ Both CA and TCA have common residual
transition formulas, see Choulakian (2006),
\begin{equation}
f_{\alpha }(i)=p_{i\ast }^{-1}\sum_{j=1}^{J}R_{\alpha -1}(i,j)u_{\alpha }(j))%
\text{ \ \ for \ \ }\alpha =1,...,k,
\end{equation}%
and%
\begin{equation}
g_{\alpha }(j)=p_{\ast j}^{-1}\sum_{i=1}^{I}R_{\alpha -1}(i,j)v_{\alpha }(i)%
\text{ \ \ for \ \ }\alpha =1,...,k,
\end{equation}%
where {\bf R}$_{\alpha }$ is the residual correspondence matrix, and ${\bf u}%
_{\alpha }$ and ${\bf v}_{\alpha }$ for \ \ $\alpha =1,...,k$ are the normed
principal axes and related to the principal coordinate scores ${\bf g}%
_{\alpha }$ and ${\bf f}_{\alpha }$ for \ \ $\alpha =1,...,k$ in the
following way. In both methods%
\begin{equation}
{\bf R}_{\alpha }={\bf P}-{\bf rc}^{^{\top }}-\sum_{\beta =1}^{\alpha }{\bf D%
}_{r}{\bf f}_{\beta }{\bf g}_{\beta }^{^{\top }}{\bf D}_{c}/\lambda _{\beta
}.
\end{equation}%
In TCA%
\begin{equation}
{\bf u}_{\alpha }=sgn({\bf g}_{\alpha })\text{ \ and \ }{\bf v}_{\alpha
}=sgn({\bf f}_{\alpha })\text{ \ \ for \ \ }\alpha =1,...,k,
\end{equation}%
so equations (14) and (15) become%
\begin{equation}
f_{\alpha }(i)=p_{i\ast }^{-1}\sum_{j=1}^{J}R_{\alpha -1}(i,j)sgn(g_{\alpha
}(j))\text{ \ \ for \ \ }\alpha =1,...,k,
\end{equation}%
and%
\begin{equation}
g_{\alpha }(j)=p_{\ast j}^{-1}\sum_{i=1}^{I}R_{\alpha -1}(i,j)sgn(f_{\alpha
}(i))\text{ \ \ for \ \ }\alpha =1,...,k.
\end{equation}%
Equations (18) and (19) help us to interpret the joint TCA maps in the
following way: $f_{\alpha }(i),$ the coordinate of point $i$ on the $\alpha $%
th axis is the signed centroid of the residual correspondence matrix within
the $p_{i\ast }^{-1}$ constant. Analogous interpretation applies to $%
g_{\alpha }(j),$ the coordinate of point $j$ on the $\alpha $th axis.
\end{itemize}

In CA%
\begin{equation}
{\bf u}_{\alpha }={\bf g}_{\alpha }/\sigma _{\alpha }\text{ \ and \ }{\bf v}%
_{\alpha }={\bf f}_{\alpha }/\sigma _{\alpha }\text{ \ \ for \ \ }\alpha
=1,...,k.
\end{equation}%
The joint interpretation of column and row categories in the CA map is based
on the well known transition formulas%
\begin{equation}
f_{\alpha }(i)=\sum_{j=1}^{J}\Pr (j|i)g_{\alpha }(j)/\sigma _{\alpha }\text{
\ \ for \ \ }\alpha =1,...,k,
\end{equation}%
and%
\begin{equation}
g_{\alpha }(j)=\sum_{i=1}^{I}\Pr (i|j)f_{\alpha }(i)/\sigma _{\alpha }\text{
\ \ for \ \ }\alpha =1,...,k,
\end{equation}%
where $\Pr (j|i)=p_{ij}/p_{i\ast },$ the conditional probability of
observing $j$ given $i$. Note that (21, 22) can be obtained from (14, 15)
via (3, 4, 20). In (21), the principal coordinate score $f_{\alpha }(i)$ is
the weighted average (centroid) of the principal coordinate scores $%
g_{\alpha }(j)$ within the $\lambda _{\alpha }^{-1}$ constant. Analogous
interpretation applies to $g_{\alpha }(j),$ the coordinate of point $j$ on
the $\alpha $th axis.

\section{Data analysis}

Here we carry out CA and TCA on three data sets mentioned in Table 1, and
comment on the remaining. The visual comparison of CA and TCA maps shows
that we can have three distinct cases: similar maps, dissimilar maps, and
partially similar maps.

\subsection{TV\ programs\ data\ set}

Table 2 presents a contingency table of size $13\times 7$ taken from Benz%
\'{e}cri (1976), where a sample of 400 individuals evaluate 13 TV programs
on a likert scale from {\it 1(excellent)} to {\it 5 (bad)}; also two other
categories of response are included {\it noopinion} on the program and {\it %
dontknow} the program. The data set is not sparse. Figure 1 displays CA and
TCA maps, where we see that both maps are similar and produce the same
interpretation. The \% of explained variation for CA (resp. for TCA) of the
first two dimensions are 70.7 \ (resp. 78\%) and 21.6 (resp (16.7); with
almost equivalent cumulative value of 92.4\% for CA and 94.7 for TCA. The
interpretation of the first two dimensions in Figure 1 will be based on two
principles: principle of dichotomy and principle of gradation.
\begin{verbatim}
4.1.1: Interpretation of the 1st axis
\end{verbatim}

We note that $C_{1}(dontknow)=700$ and $SC_{1}(dontknow)=-500$ as given at
the bottom of Table 2; so the first axis represents the dichotomy between
ignorance and knowledge: where the response category {\it dontknow} opposes
to the 5 likert response scales; {\it noopinion} is near the origin.
\begin{verbatim}
4.1.2: Interpretation of the 2nd axis
\end{verbatim}

The 5 Likert response categories are ordered from {\it excellent} to {\it bad%
}.
\begin{verbatim}
4.1.3: Interpretation of the TV programs
\end{verbatim}

Programs 2 and 12 are considered {\it excellent} and {\it verygood};
programs 10, 11 and 9 are mostly {\it unknown,} and so on{\it .}

It is important to note that, the response category {\it dontknow} is very
influential in both methods CA and TCA, and it reveals a central important
feature of the data: in TCA, the category $dontknow$ contributes only to the
first axis, because it attains the maximum value of its contribution, $%
|SC|=500$, see equation (9); but this is not the case in CA.

\subsection{Rodent\ species\ abundance\ data\ set}

Table 2 displays abundance data ({\bf N}) of size $28\times 9$ (equivalent
to {\bf M} of size $21\times 9)$, where 9 species of rodents have been
counted at each of 28 sites in California. For the interested reader, we
identify the 9 rodents by their scientific names: {\it rod1=Rt.rattus,
rod2=Mus.musculus, rod3=Pm.californicus, rod4=Pm.eremicus,
rod5=Rs.megalotis, rod6=N.fuscipes, rod7\ =N.lepida},\\{\it rod8=Pg.fallax, }and%
{\it \ rod9=M.californicus}. Genus abbreviations are: {\it Rt} (Rattus),
{\it Rs} (Reithrodontomys), {\it Mus} (Mus), {\it Pm} (Peromyscus), {\it Pg}
(Perognathus), {\it N} (Neotoma) and {\it M} (Microtus). Rattus and Mus,
rodents 1 and 2, are invasive species, whereas the others are native. This
data set is very interesting, because we see that it has, in particular,
three specificities which characterize our concept of sparsity: rare
observations, a zero-block structure and relatively high-valued cells. It is
sparse based on the 7-number summary calculated in Table 1. It was proposed
in 2014 as an exercise in a course on an ecology workshop in UBC in Canada;
the workshop site mentions that the data set is downloaded from the web site
of Quinn and Keough (2002), and it can be found at

https://www.zoology.ubc.ca/\symbol{126}%
bio501/R/workshops/workshops-\\multivariate-methods/

The instructor of the course suggested the analysis of this data set by CA
in two rounds:

In round 1 the invasive species dominate, see the CA map in Figure 2. This
fact is confirmed by looking at the contibutions in Table 2, where we find $%
C_{1}(rod2)=750,$ so the first axis in the CA map is dominated by rodent 2; $%
C_{2}(rod1)=854$, so the second axis is dominated by rodent 1. The
highlighted subset of sites, 7-11, 14-17, 21-22, 24-25, which are completely
associated only with the invasive rodents 1 and 2, characterizes the CA
solution; their total weight is $84/1002=8.38\%.$ The minimal matrix ${\bf M}
$ has a zero-block structure and it can be reexpressed as%
\[
{\bf M}=\left(
\begin{array}{cc}
{\bf M}_{1} & {\bf M}_{2} \\
{\bf M}_{3} & {\bf 0}%
\end{array}%
\right) ,
\]%
where
\[
{\bf M}_{3}=\left(
\begin{array}{cc}
0 & 59 \\
3 & 8 \\
1 & 2 \\
1 & 3 \\
4 & 0 \\
2 & 1%
\end{array}%
\right) ,
\]%
and the submatrix $(%
\begin{array}{cc}
{\bf M}_{3} & {\bf 0}%
\end{array}%
)$ represents the 13 highlighted sites. We conclude that the combination of
the high count cell 59 (representing 7 rare sites), the last five rows in $%
{\bf M}_{3}\ $(representing 6 rare sites), and the large zero-block in ${\bf %
M}$, created this particular CA solution.

In round 2 the instructor suggested eliminating rodents 1 and 2 and their
associated sites, and carrying out a second application of CA on the reduced
data set representing only the native species (the CA map is not shown).

Figure 2 displays the principal maps produced by CA and TCA, where the two
invasive species and their associated sites are fenced by linear segments:
they are completely different. It is evident that for this data the TCA map
is much more informative than the corresponding CA map; further, one TCA map
is as informative as two CA maps obtained from the two rounds.
\begin{verbatim}
4.2.1  Interpretation of the TCA map
\end{verbatim}

Let us interpret the TCA biplot, the lower diagram in Figure 2. We note that
the two invasive species are grouped and found in the first quadrant of the
TCA map; further, they are associated with the 13 highlighted subset of
sites that we enumerated above. The contibutions to dimensions 1 and 2 of
the rodents, $SC_{1}$ and $SC_{2}$ displayed in Table 2, show that the TCA
map is dominated by the four most frequent species: rodents 2, 6, 3 and 4;
and each of them occupy a quadrant in the TCA map.

Let us interpret the most frequent ($freq$) species rodent 3 ( $freq=467$
out of $1002$): it dominates the third quadrant, and it is associated with
site3 ($freq=36$), site5 ($freq=63$), site13 ($freq=39$), site18 ($freq=78$%
), site27 ($freq=29$) and site28 ($freq=10$). Rod3 is also associated with
rod5, their positions are quite near on Figure 2; looking at the entries in
the column of rod5, we see that its high frequency sites are site27 ($%
freq=10 $), site18 ($freq=10$), site6 ($freq=12$), site5 ($freq=11$);
further these high frequency sites 27, 18, 6 and 5 also characterize rod3.
We also note that site6 is associated with both species rod3 ($freq=48$) and
rod4 ($freq=35$), and its position is in between rod3 and rod4; however it
is found in quadrant 3, because $(35/125)>48/467$.
\begin{verbatim}
4.2.2 Comparison
\end{verbatim}

By comparing the first two principal dimensions in CA and TCA, we note the
following two facts:

a) Rodent 2, with nonnegligeable weight (around 10\%), has a very high
influence on the first principal axis in CA ($C_{1}(rod2)=750);$ but it
distributes its influence onto the first two principal axes in TCA ($%
SC_{1}(rod2)=-196$ and $SC_{2}(rod2)=-238)$.

b) Rodent 1 is a rare influential point in CA ($weight=1.4\%$ and \\$%
C_{2}(rod1)=854$); but is no more influential in TCA \\($SC_{1}(rod1)=-23$ and
$SC_{2}(rod1)=-26)$.

We conclude that in Figure 2, the CA map emphasized some particular aspects
of the data set; while the TCA map revealed the central abundances in Table
2.

\subsection{Macro abundance data set}

This macroinvertebrate sparse abundance data set of size $197\times 40$ was
also considered by Greenacre (2013). Figures 3 and 4 display the CA and TCA
maps: The first dimension has almost the same separation of the 40 sites, so
the same interpretation; while the second dimension seems somewhat
different. For this reason we labeled the map similarity partial in Table 1.

\subsection{Remaining data sets}

Here, we just give references for the remaining data sets listed in Table 1.

The five ecology abundance data sets numbered 3 to 7 are available in
Greenacre (2013) and he discussed them in his essay.

Mallet-Gauthier and Choulakian (2015) analyse {\it Punta Milazesse} and {\it %
Iverjsford} abundance data in archeolgy; they reproduce the data sets,
provide CA and TCA maps and their interpretations.

Choulakian et al. (2006) is the reference for the {\it Synoptic Gospels}
textual count data; they provide CA and TCA maps and discuss their stability.

\section{Sparsest contingency tables}

Sparsest contingency tables are diagonal contingency tables, as we defined
in section 2. Here, we show that CA and TCA results are completely different.

\subsection{CA of sparsest contingency tables}

Let ${\bf N}$ be a diagonal contingency table of size ${\it I}$, then it is
well known, see for instance Benz\'{e}cri (1973, p. 188) that CA of ${\bf N}$%
{\bf \ }produces\ $(I-1)$ dispersion measures of 1; that is, $\sigma
_{\alpha }^{CA}=1$ for $\alpha =1,...,I-1.$ So, CA shows that ${\bf N}${\bf %
\ }is composed of $I$ diagonal blocks. A similar result is known in spectral
clustering as Fiedler's theorem, see Choulakian and de Tibeiro (2013).

\subsection{TCA of sparsest contingency tables}

Let ${\bf P}_{{\bf N}}=diag(p_{1},...,p_{I})$ be the correspondence matrix
of a diagonal contingency table ${\bf N.}$ Then we have the following easily
proven result:

{\bf Corollary to Lemma 2}:{\bf \ }$\sigma _{1}^{TCA}=1$ if and only if
there is a subset $S\subset \left\{ 1,...,I\right\} $ such that $\sum_{i\in
S}p_{i}=0.5.$

{\it proof}: By Lemma 2,
\begin{eqnarray*}
\sigma _{1}^{TCA} &=&4\ {\bf v}_{1+}^{\prime }{\bf R}_{0}{\bf u}_{1+} \\
&=&4\ {\bf u}_{1+}^{\prime }{\bf R}_{0}{\bf u}_{1+}\text{, \ for }{\bf R}_{0}%
\text{ is symmetric} \\
&=&4\ \sum_{i\in S}p_{i}(1-\sum_{i\in S}p_{i}),
\end{eqnarray*}%
where $S=\left\{ i:{\bf u}_{1+}(i)=1\text{ for }1,...,I\right\} $, and the
required result follows.

\subsection{Examples}

We present three exemples, two contrived and one real.
\begin{verbatim}
5.3.1. Example 1
\end{verbatim}

Let ${\bf N=Diag(}1,\ 2,\ 3,\ 4,\ 6);$ then CA produces identical singular
values $\sigma _{1}^{CA}=\sigma _{2}^{CA}=\sigma _{3}^{CA}=\sigma
_{4}^{CA}=\ 1;$ while TCA produces\ $\sigma _{1}^{TCA}=1$ for $2+6=1+3+4,\ $%
and the remaining dispersion measures are $$\sigma _{\alpha }^{TCA}=0.875,\
0.85714\ \ \ $$ and$\ \ 0.18750.$
\begin{verbatim}
5.3.2. Example 2
\end{verbatim}

\ Let ${\bf N=Diag(}1,\ 2,\ 3,\ 4,\ 5);$ then CA produces $\sigma
_{1}^{CA}=\sigma _{2}^{CA}=\sigma _{3}^{CA}=\sigma _{4}^{CA}=\ 1;$ while TCA
produces\ dispersion measures $$\sigma _{\alpha }^{TCA}=0.99556,\ 0.95714,\
0.95522\ \ \mbox{and}\ \ 0.17778.$$
\begin{verbatim}
5.3.3. Texel abundance data set
\end{verbatim}

Greenacre (2013) described this data set of size $285\times 220$ "as large
and very sparse table of vegetation abundances on a coastal sand dune area
on the island of Texel, the Netherlands". CA is of no help for the analysis
of this table, because according to Greenacre CA needs as much as 71
dimensions. This is evident by looking at the sequence of CA singular
values: The first five singular values are: $\sigma _{\alpha }^{CA}=\ 1,$\ $%
0.9932,$\ $0.9908,$\ $0.9798$ and $0.9761;$ $\sigma _{\alpha }^{CA}=\ 1$
means that by permuting rows and columns of the data set, the data can be
represented in two diagonal blocks. The corresponding TCA dispersion
measures are: $\sigma _{\alpha }^{TCA}=0.8026,\ 0.7768,\ 0.7331,\ 0.7106\ $%
and $0.7012$. Figures 5 and 6 display the CA and TCA maps, which are
completely different. We leave the interpretation of the TCA map, if there
is any, to the ecologists.

\section{Conclusion}

The fundamental aim of CA\ and TCA is to produce interpretable maps that
reflect central contents in a data set. In this paper, first we provided a
7-number quantification of sparsity; then we showed that for sparse
contingency tables CA and TCA maps can differ with positive probability,
because a map produced by CA or TCA is dependent on the underlying geometry,
Euclidean or Taxicab. Based on our experience, we suggest the analysis of a
data set by both methods CA and TCA: Like a cubist painting where an object
is painted from different angles, sometimes the views are similar, and at
other times dissimilar or partially similar.
\begin{verbatim}
Acknowledgements.
\end{verbatim}

This research has been supported by NSERC of Canada. The detailed comments
of M. Greenacre and two anonymous reviewers improved the presentation of
this essay significantly.\bigskip

Appendix

{\bf Lemma 1}: $\%(0\in {\bf N})\leq 100(1-\frac{1}{\min (I,J)}).$

The proof is very easy by considering 2 distinct cases of data sets, square $%
(I=J)$ and rectangular $(J>I)$.

Case 1: If ${\bf M}$ is diagonal and has exactly $I$ nonzero cells, then by
permuting some rows and columns, it can be rearranged into a diagonal
contingency table; thus
\begin{eqnarray*}
\%(0 &\in &{\bf M})=100(I^{2}-I)/I^{2} \\
&=&100(1-1/I),
\end{eqnarray*}%
which is the upper bound. If ${\bf M}$ is a square contingency table but not
diagonal, by permuting some rows and columns, it can be rearranged into a
square contingency table with all diagonal cells nonzero plus some, say, $%
\alpha $ number of nondiagonal nonzero cells. Then it is evident that
\begin{eqnarray*}
\%(0 &\in &{\bf M})=100(I^{2}-I-\alpha )/I^{2} \\
&\leq &100(1-1/I).
\end{eqnarray*}

Case 2: ${\bf M}$ is rectangular and $(J>I)$. Then ${\bf M}=({\bf M}_{1}|%
{\bf M}_{2})$, where ${\bf M}_{1}$ is square with nonzero diagonal elements
of size $I\times I$ and has $\alpha $ number of nondiagonal nonzero cells ; $%
{\bf M}_{2}$ is rectangular of size $I\times (J-I),$ such that each column
of ${\bf M}_{2}$ has exactly $\beta _{i}$ nonzero cells for $i=1,...,(J-I)$
and $2\leq $ $\beta _{i}\leq I.\ $Then%
\begin{eqnarray*}
\%(0 &\in &{\bf M})=\frac{100\left[ (IJ-I-\alpha -\sum_{i=1}^{J-I}\beta _{i})%
\right] }{IJ} \\
&\leq &100(1-1/I),
\end{eqnarray*}%
because $\sum_{i=1}^{J-I}\beta _{i}\geq 2(J-I).\bigskip $

{\bf References}\medskip

Agresti, A. (2002). {\it Categorical Data Analysis}. New Jersey: John Wiley
\& Sons, 2nd edition.

Agresti, A. and Yang, M.C. (1987). An empirical investigation of some
effects of sparseness in contingency tables. {\it Computational Statistics
\& Data Analysis}, 5, 9--21.

Beh, E. and Lombardo, R. (2014). {\it Correspondence Analysis: Theory,
Practice and New Strategies}. N.Y: Wiley.

Benz\'{e}cri, J.P. (1976). Sur le codage r\'{e}duit d'un vecteur de
description en analyse des correspondances. {\it Les Cahiers de l'analyse
des donn\'{e}es}, 1(2), 127-136.

Benz\'{e}cri, J.P. (1973).\ {\it L'Analyse des Donn\'{e}es: Vol. 2:
L'Analyse des Correspondances}. Paris: Dunod.

Benz\'{e}cri, J.P (1992). {\it Correspondence Analysis Handbook}. N.Y:
Marcel Dekker.

Choulakian, V. (2006). Taxicab correspondence analysis. {\it Psychometrika,}
71, 333-345

Choulakian, V. (2008). Taxicab correspondence analysis of contingency tables
with one heavyweight column. {\it Psychometrika}, 73, 309-319.

Choulakian, V., Kasparian, S., Miyake, M., Akama, H., Makoshi, N., Nakagawa,
M. (2006). A statistical analysis of synoptic gospels. {\it JADT'2006}, pp.
281-288.

Choulakian, V. and de Tibeiro, J. (2013). Graph partitioning by
correspondence analysis and taxicab correspondence analysis. {\it Journal of
Classification}, 30, 397-427.

Choulakian, V., Allard, J. and Simonetti, B. (2013). Multiple taxicab
correspondence analysis of a survey related to health services. {\it Journal
of Data Science}, 11(2), 205-229.

Choulakian, V., Simonetti, B. and Gia, T.P. (2014). Some further aspects of
taxicab correspondence analysis. {\it Statistical Methods and Applications},
available online.

Choulakian, V. (2016). Matrix factorizations based on induced norms.\\ {\it
Statistics, Optimization and Information Computing}, 4, 1-14.

Gifi, A. (1990). {\it Nonlinear Multivariate Analysis. }N.Y:{\it \ } Wiley.

Greenacre, M. (1984). {\it Theory and Applications of Correspondence Analysis%
}. Academic Press, London.

Greenacre, M. (2013). The contributions of rare objects in correspondence
analysis. {\it Ecology}, 94(1), 241-249.

Khot, S. ans Naor, A. (2012). Grothendieck-type inequalities in
combinatorial optimization. {\it Communications in Pure and Applied
Mathematics}, 65 (7), 992-1035.

Kraus, K. (2012). {\it On the Measurement of Model Fit for Sparse
Categorical Data}. Ph.D thesis, Acta Universitatis Upsaliensis, Uppsala.

Le Roux, B. and Rouanet, H. (2004). {\it Geometric Data Analysis. From
Correspondence Analysis to Structured Data Analysis}. Dordrecht:
Kluwer--Springer.

Mallet-Gauthier, S. and Choulakian, V. (2015). Taxicab correspondence
analysis of abundance data in archeology: three case studies revisited. {\it %
Archeologia e Calcolatori}, 26, 77-94.

Murtagh, F. (2005). {\it Correspondence Analysis and Data Coding with Java
and R}. Boca Raton, FL., Chapman \& Hall/CRC.

Nishisato, S. (1984). Forced classification: A simple application of a
quantification method. {\it Psychometrika}, 49(1), 25-36.

Nishisato, S. (2007). {\it Mutidimensional nonlinear descriptive analysis}.
Chapman \& Hall/CRC, Boca Raton.

Nishisato, S. (1998). Graphing is believing: interpretable graphs for dual
scaling. In Blasius, J. and Greenacre, M. (eds.), "{\it Visualization of
Categorical Data}", Academic Press, NY, 185-196.

Nowak, E. and Bar-Hen, A. (2005). Influence function and correspondence
analysis. {\it Journal of Statistical Planning and Inference}, 134, 26-35.

Pisier, G. (2012). Grothendieck's theorem, past and present. {\it Bulletin
of the American Mathematical Society}, 49 (2), 237-323.

Quinn, G. and Keough, M. (2002). {\it Experimental Design and Data Analysis
for Biologists}. Cambridge Univ. Press, Cambridge, UK.

Radavicius, M. and Samusenko, P. (2012). Goodness-of-fit tests for sparse
nominal data based on grouping. {\it Nonlinear Analysis: Modelling and
Control}, 17 (4), 489--501.

Rao, C.R. (1995). A review of canonical coordinates and an alternative to
correspondence analysis. {\it Q\"{u}estii\'{o}}, 19, 23-63.

Schlick, Th. (2000). {\it Readings in the philosophy of Science: from
positivism to post modernism}.\\
 Mayfield Publishing Company, Mountain View, California.

Tukey, J. (1977). {\it Exploratory Data Analysis}.\\ Addison-Wesley,
Massachusetts.\\
\newpage
{\tiny\begin{tabular}{|l|l|l|l|l|l|l|l|}
\multicolumn{8}{l|}{\bf Table 2: TV progams data.} \\ \hline
programs & excellent & verygood & good & average & bad & noopinion & dontknow
\\ \hline
1 & 9 & 28 & 89 & 124 & 51 & 19 & 71 \\ \hline
2 & 31 & 87 & 165 & 63 & 24 & 4 & 17 \\ \hline
3 & 7 & 21 & 65 & 103 & 83 & 8 & 103 \\ \hline
4 & 3 & 26 & 121 & 142 & 45 & 11 & 43 \\ \hline
5 & 17 & 40 & 117 & 111 & 83 & 16 & 7 \\ \hline
6 & 8 & 35 & 115 & 119 & 78 & 6 & 28 \\ \hline
7 & 4 & 22 & 73 & 56 & 77 & 12 & 147 \\ \hline
8 & 15 & 44 & 102 & 83 & 32 & 25 & 90 \\ \hline
9 & 5 & 18 & 63 & 61 & 15 & 9 & 219 \\ \hline
10 & 8 & 15 & 40 & 37 & 8 & 12 & 271 \\ \hline
11 & 5 & 16 & 64 & 54 & 15 & 17 & 220 \\ \hline
12 & 29 & 87 & 140 & 62 & 24 & 9 & 40 \\ \hline
13 & 12 & 18 & 89 & 95 & 41 & 9 & 127 \\ \hline\hline
total & 153 & 457 & 1243 & 1110 & 576 & 157 & 1383 \\ \hline\hline
$C_{1}$ & 24 & 83 & 106 & 45 & 40 & 1 & {\bf 700} \\ \hline
\multicolumn{1}{|l|}{$C_{2}$} & 128 & 285 & 63 & 181 & 330 & 2 & 11 \\
\hline\hline
$SC_{1}$ & -28 & -96 & -165 & -137 & -73 & -2 & {\bf 500} \\ \hline
$SC_{2}$ & -82 & -235 & -173 & 222 & 278 & -10 & 0 \\ \hline\hline
\end{tabular}
}\\
\bigskip
\begin{tabular}{|l|l|l|l|l|l|l|l|l|l|}
\multicolumn{10}{l|}{\bf Table 3: Rodent species abundance data.} \\ \hline
{\it Sites} & rod1 & rod2 & rod3 & rod4 & rod5 & rod6 & rod7 & rod8 & rod9
\\ \hline
1 &  & 13 & 3 & 1 & 1 & 2 &  &  &  \\ \hline
2 &  & 1 & 57 & 65 & 9 & 16 & 8 & 2 & 3 \\ \hline
3 &  & 4 & 36 &  & 2 & 9 &  &  &  \\ \hline
4 &  & 4 & 53 & 1 & 5 & 30 &  & 18 & 3 \\ \hline
5 &  & 2 & 63 & 21 & 11 & 16 &  &  &  \\ \hline
6 &  & 1 & 48 & 35 & 12 & 8 & 12 & 2 & 2 \\ \hline
{\bf 7} &  & {\bf 11} &  &  &  &  &  &  &  \\ \hline
{\bf 8} &  & {\bf 16} &  &  &  &  &  &  &  \\ \hline
{\bf 9} & {\bf 3} & {\bf 8} &  &  &  &  &  &  &  \\ \hline
{\bf 10} & {\bf 1} & {\bf 2} &  &  &  &  &  &  &  \\ \hline
{\bf 11} &  & {\bf 9} &  &  &  &  &  &  &  \\ \hline
12 &  & 3 & 1 &  & 5 & 16 &  & 7 &  \\ \hline
13 &  & 4 & 39 &  & 4 & 12 &  &  &  \\ \hline
{\bf 14} & {\bf 1} & {\bf 3} &  &  &  &  &  &  &  \\ \hline
{\bf 15} &  & {\bf 11} &  &  &  &  &  &  &  \\ \hline
{\bf 16} &  & {\bf 4} &  &  &  &  &  &  &  \\ \hline
{\bf 17} & {\bf 3} &  &  &  &  &  &  &  &  \\ \hline
18 &  & 2 & 78 &  & 10 & 14 &  & 4 &  \\ \hline
19 &  &  & 1 &  &  &  &  &  &  \\ \hline
20 & 3 &  & 27 & 1 &  &  &  &  &  \\ \hline
{\bf 21} & {\bf 2} & {\bf 1} &  &  &  &  &  &  &  \\ \hline
{\bf 22} &  & {\bf 3} &  &  &  &  &  &  &  \\ \hline
23 &  &  &  &  & 2 & 8 &  & 2 &  \\ \hline
{\bf 24} & {\bf 1} &  &  &  &  &  &  &  &  \\ \hline
{\bf 25} &  & {\bf 5} &  &  &  &  &  &  &  \\ \hline
26 &  &  & 22 &  &  & 11 &  & 2 &  \\ \hline
27 &  &  & 29 &  & 10 & 9 &  & 1 &  \\ \hline
28 &  &  & 10 & 1 &  & 1 &  &  &  \\ \hline\hline
total & {\it 14} & {\bf 107} & {\bf 467} & {\bf 125} & 71 & {\bf 152} & 20 &
38 & 8 \\ \hline\hline
$\sigma _{\alpha }^{TCA}$ & 0.478 & 0.422 & 0.347 & 0.138 & 0.120 & 0.091 &
0.061 & 0.010 &  \\ \hline\hline
$\sigma _{\alpha }^{CA}$ & 0.864 & 0.678 & 0.536 & 0.391 & 0.189 & 0.157 &
0.107 & 0.045 &  \\ \hline\hline
$C_{1}$ & 127 & {\bf 750} & 59 & 29 & 9 & 15 & 5 & 4 & 2 \\ \hline
$C_{2}$ & {\bf 854} & 140 & 3 & 0 & 0 & 2 & 0 & 1 & 0 \\ \hline\hline
$SC_{1}$ & -23 & {\bf -196} & {\bf 298} & {\bf -221} & 22 & {\bf 135} & -51
& 44 & -8 \\ \hline
$SC_{2}$ & -26 & {\bf -238} & {\bf 202} & {\bf 224} & 32 & {\bf -139} & 42 &
-95 & -1 \\ \hline
\end{tabular}

\ \ \ \ \ \ \
\begin{figure}[ptb]
\begin{center}
\includegraphics[width=5.2in]{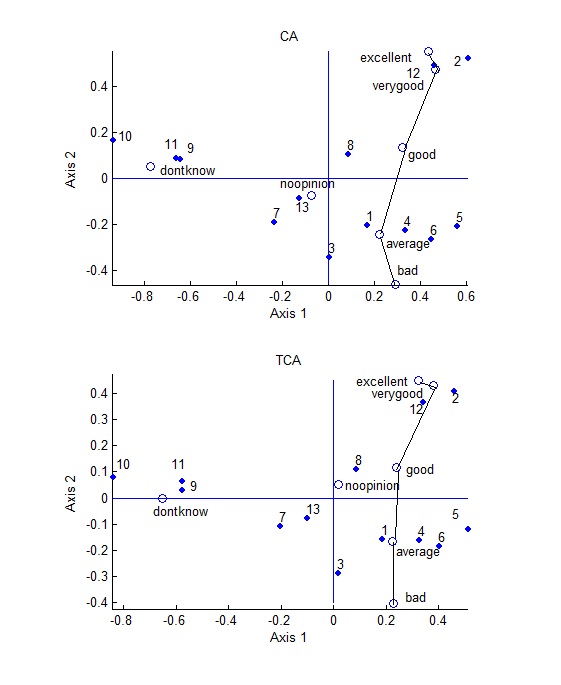}\\
\caption{CA and TCA biplots of TV Programs data.}
\end{center}
\end{figure}%
\begin{figure}[ptb]
\begin{center}
\includegraphics[width=5.3in]{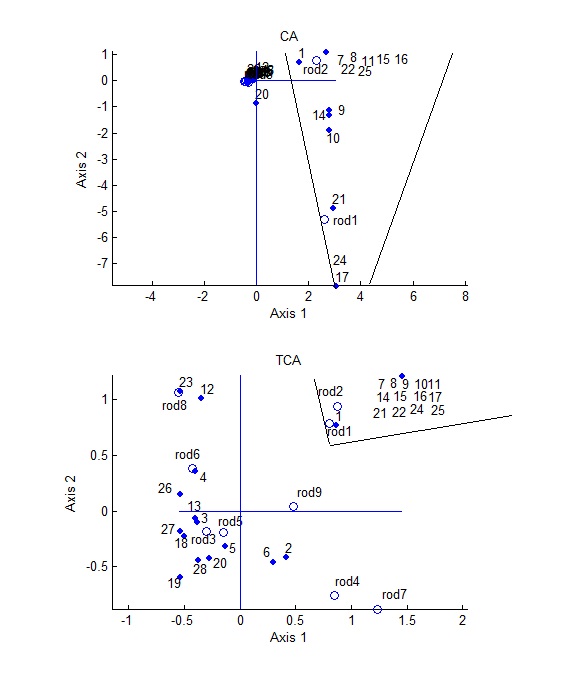}\\
\caption{CA and TCA biplots of Rodents data.}
\end{center}
\end{figure}
\begin{figure}[ptb]
\begin{center}
\includegraphics[width=4.3in]{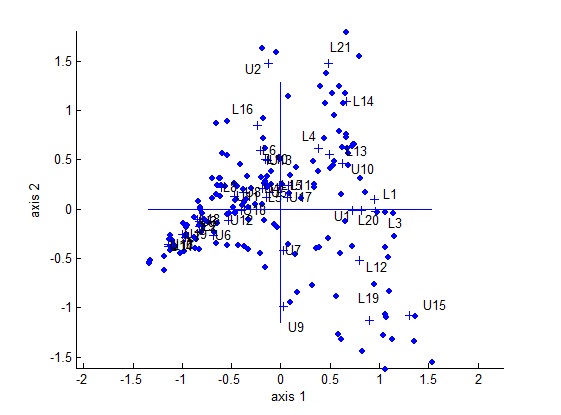}\\
\caption{CA map of Macro data.}
\end{center}
\end{figure}%
\begin{figure}[ptb]
\begin{center}
\includegraphics[width=4.3in]{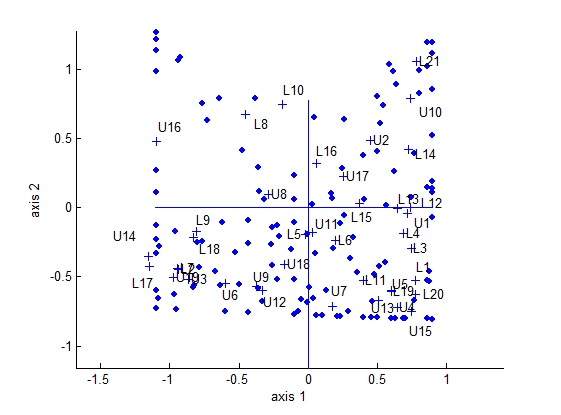}\\
\caption{TCA map of Macro data.}
\end{center}
\end{figure}%
\begin{figure}[ptb]
\begin{center}
\includegraphics[width=4.3in]{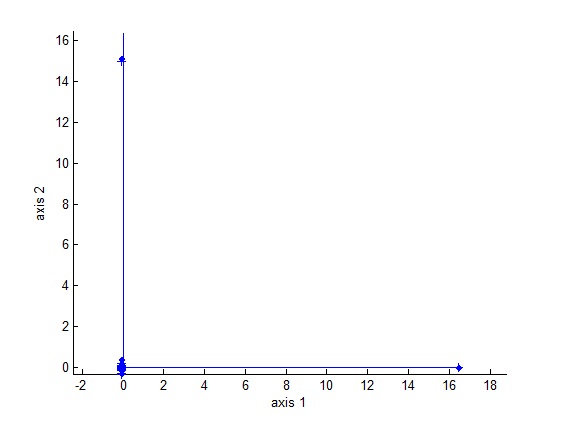}\\
\caption{CA map of Texel data.}
\end{center}
\end{figure}%
\begin{figure}[ptb]\begin{center}
\includegraphics[width=4.3in]{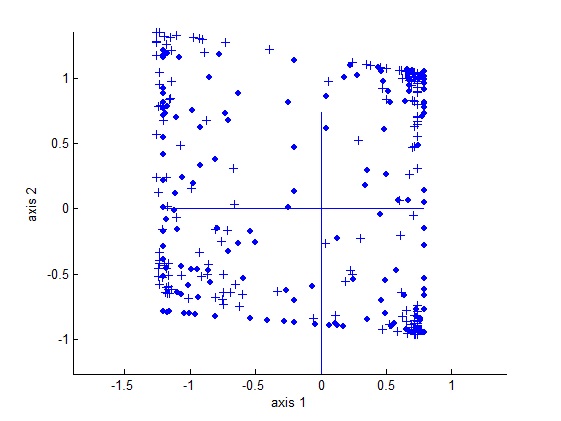}\\
\caption{TCA map of Texel data.}
\end{center}
\end{figure}%
\end{document}